# Crystal structure and electronic structure of $CePt_2In_7$

T. Klimczuk[1], O. Walter[2], L. Müchler[3], J.W. Krizan[3], F. Kinnart[2] and R. J. Cava[1]

[1]*Faculty of Applied Physics, Gdansk University of Technology, Narutowicza 11/12, 80-233 Gdansk, Poland*

[2]*European Commission, Joint Research Center, Institute for Transuranium Elements, Postfach 2340, Karlsruhe, D-76125 Germany*

[3]*Department of Chemistry, Princeton University, Princeton, NJ 08544, USA*

**Abstract**

We report a corrected crystal structure for the $CePt_2In_7$ superconductor, refined from single crystal x-ray diffraction data. The corrected crystal structure shows a different Pt-In stacking along the *c*-direction in this layered material than was previously reported. In addition, all the atomic sites are fully occupied with no evidence of atom site mixing, resolving a discrepancy between the observed high resistivity ratio of the material and the atomic disorder present in the previous structural model  The Ce-Pt distance and coordination is typical of that seen in all other reported $Ce_nM_mIn_{3n+2m}$ compounds. Our band structure calculations based on the correct structure reveal three bands at the Fermi level that are more three dimensional than those previously proposed, and Density functional theory (DFT) calculations show that the new structure has a significantly lower energy.

**Introduction**

The first superconductor whose detailed phenomenology could not be described by conventional BCS theory, $CeCu_2Si_2$, was reported by Steglich and coworkers three decades ago. $CeCu_2Si_2$ forms in the "122 type" $ThCr_2Si_2$ crystal structure and superconducts below ~ 0.5 K.[1] This was unexpected, as it was found only four years after $CeAl_3$ was reported as the first heavy-fermion compound.[2] A very important class of heavy-fermion superconducting materials of current interest, the "115 superconductors" $CeMIn_5$ (M=Co, Rh, Ir), was announced in 2001.[3] The significance of the "Ce-115" family originates from the unique opportunity it offers to explore the relationship between unconventional superconductivity and magnetism, and its study has resulted in more than a thousand papers published in the last decade. It has been shown that "Ce-115" compounds belong to a more general $A_nM_mIn_{3n+2m}$ family, where M is a group 9 transition metal and A is not only Ce but can also be U, Np or Pu. Recently it has been shown that extension to group 10 metals is possible, but only for a higher order members of the structural family, i.e. n=2, m=1 ($Ce_2PdIn_8$ – refs. 4, 5, 6), n=1, m=2 ($CePt_2In_7$ – ref. 7), n=3, m=1 ($Ce_3PdIn_{11}$ – ref. 8 and $Ce_3PtIn_{11}$ – ref. 9). The current understanding of the heavy-fermion superconductivity in the $A_nM_mIn_{3n+2m}$ family is discussed in ref. 10.

The crystal structures for the four known members of $A_nM_mIn_{3n+2m}$ family are presented in Figure 1. The first member ($CeIn_3$) forms in a primitive cubic structure, and is a common building unit for the more complex $A_nM_mIn_{3n+2m}$ compounds. The next two members, $CeMIn_5$ and $Ce_2MIn_8$, form in a primitive tetragonal structure (P4/mmm), whereas $CePt_2In_7$ crystallizes in a body-centered tetragonal structure (I4/mmm). These higher order members consist of $CeIn_3$ and $MIn_2$ blocks stacked along the *c*-axis. As can be seen in Figure 1, the largest ($c$ = 21.57 Å) unit cell is observed for $CePt_2In_7$. The crystal structure of this compound was first described by Kurenbaeva et. al. (ref. 7) and later confirmed by Tobash et al. (ref. 11). Here we show that that model is not correct, and present the correct model for the crystal structure of $CePt_2In_7$ (Fig. 1 d), employed in the interpretation of NMR data on this compound.[12] Electronic structure calculations are performed based on this crystal structure, and the consequences of the difference in structure are described.

**Experimental**

Plate-like single crystals of CePt$_2$In$_7$ were grown from a high-temperature solution as described in ref. 11. The starting high purity (99.9% Pt, 99.9% Ce, 99.99% In) metals, in the ratio of Ce:Pt:In 1:3:20, were placed in an alumina crucible and sealed under vacuum in a quartz tube. The tube was heated to 1180 °C and kept at that temperature for 8 h, and then cooled to 800 °C over a period of 1-1.5 hour, followed by slow cooling to 400°C at the rate of 4 °C/h. Excess In was spun off at 400 °C with the aid of a centrifuge.

Single crystal XRD measurements were performed using a Bruker APEX II Quazar diffractometer with monochromated Mo K$_\alpha$ irradiation collecting four spheres of data in the θ range from 3.78 to 38.54°. Frames were collected with an irradiation time of 4 s per frame and combined ω- and φ- scan technique with $\Delta\omega = \Delta\varphi = 0.5°$. Data were corrected to Lorentz and polarization effects, and an experimental adsorption correction with SADABS [13] was applied. The structures were solved by direct methods and refined to an optimum R1 value with SHELX-2013. [14]

The electronic structure calculations were performed in the framework of density functional theory (DFT) using the WIEN2K code with a full-potential linearized augmented plane-wave and local orbitals [FP-LAPW+lo] basis [15] together with the Perdew-Burke-Ernzerhof (PBE) parametrization [16] of the generalized gradient approximation (GGA) as the exchange-correlation functional. The plane-wave cutoff parameter RKMAX was set to 8 and the Brillouin zone was sampled by 10,000 *k* points. The results of calculations that do not include spin-orbit coupling (SOC) are presented. The inclusion of SOC did not change significantly the calculated total energy for the different structural models.

**Results**

Single crystal x-ray diffraction patterns were collected at a temperature of 173 K. The structure was solved using direct methods on F$^2$. Three structural models were tested (Table 2) the and the refined lattice parameters are the same for all models. They are approximately 0.2% larger and smaller than those reported in reference 7 and 11, respectively For the first two (A and B), the atoms have the same positional coordinates but the occupations of the sites are taken as variable. Model A assumes 100% site occupation and has no atom admixture present, i.e. it has no structural disorder. This model, a fully ordered variant of the previously reported structure, is certainly not correct as it returns a negative thermal parameter ($U_{eq}$) for the In1 position, a large thermal parameter for the Ce position and an unacceptable goodness of fit value (GOF=2.143). Model B has the same atomic positions as Model A and is based on the crystal structure reported in ref. 7 and confirmed in ref. 11. This model assumes partial Ce site occupancy (85%), and, in addition, that the In1 site is an admixture of In and Pt in the ratio 75:25. The refinements of model B give a much lower GOF= 1.089. However, as the authors of ref. 11 pointed out, that presence of a large number of Ce site vacancies, and the strong structural disorder due to Pt/In site mixing is in contradiction to the observed very low residual resistivity ($\rho_0 = 0.2$ $\mu\Omega$ cm) and very high residual resistivity ratio (RRR = 400) observed for this material.

We have corrected this structural model by replacing the Indium in the Wyckoff *2b* position by Cerium. The removed indium is then placed in the Wyckoff *2a* position (see Table 2), maintaining the material stoichiometry. The structure has 100% site occupancies and no structural disorder. The quality of the refinement is excellent. The GOF obtained is 1.022, and the residual electron densities are excellent. This structure, with no structural disorder, is, as opposed to the former one, reconciles the discrepancy between the transport properties and crystal structure of the compound.

In the new crystal structure (Fig. 1d), the stacking of the Pt-In blocks is changed relative to the previously reported model. The most important consequence of this difference is a much shorter Ce-Pt distance. The inset of Figure 1 presents the local coordination environment of a Ce atom (green ball), inside a polyhedron formed by In atoms (purple balls). The corrected model places the Pt in a position where it caps the Ce-In polyhedra giving a Ce-Pt distance of $d_{Ce-Pt}$ =3.76 Å, as opposed to the former model where Pt is in a position where it sits over a corner of the polyhedra ($d_{Ce-Pt}$ = 4.96 Å). The Ce-Pt environment and separation in the correct structural model is in a very good agreement with $d_{Ce-Pt}$ observed for all other Ce-based $Ce_nM_mIn_{3n+2m}$ compounds, an additional indication of its correctness.

Figure 2 shows the calculated density of states (DOS) for the ordered models: former (A) and correct (C). (The density of states for the highly disordered model (B) cannot reliably be calculated by the DFT method at the present time). The DOS at the Fermi level is similar, but 2 eV above and below the Fermi level there are significant differences, for example the presence of a small shoulder slightly above the Fermi level that is due to Ce-f states. Despite a similar DOS, the Fermi surfaces derived from the structures are quite different, as is shown in Figure 3. For both structures, there are 4 bands crossing the Fermi level. For the former model (A), there are 3 cylinders at the edges of the Brillouin zone with little $k_z$ dispersion, together with a more delocalized band that also shows little dispersion in the $k_z$ direction; this can be attributed to the more 2-dimensional character of the structure of model A. The calculated Fermi surface is in agreement with that reported in ref. 17. For the correct model (C), there are only two bands at the Brillouin zone boundary, one with cylindrical symmetry and one with more rectangular symmetry. Both bands have 2-dimensional character, however both are slightly warped. One of the 4 bands is almost fully occupied and only provides small pockets around the zone boundary in the $k_z$ direction. The last band is more 3-dimensional, with pockets at the zone boundaries and a pocket around the Γ-point. Using the experimental structures, the corrected structural model calculates to be lower in energy by 1.148 eV / cell compared to model A. This is a significant difference, helping to explain the stability of our corrected structure. (The introduction of spin-orbit coupling (SOC) into the calculation does not influence the states at the Fermi level) . The Ce-*f*-states are splitwith a large peak in the DOS approximately 0.2 eV above the Fermi level. This 4*f*-peak is also observed in $CeRhIn_5$, where it coincides with the Fermi level (ref. 17).

**Conclusions**

Knowledge of the crystal structure of a material with interesting physical properties is critical for its understanding. The structural parameters are, for example, required as the input for electronic structure calculations and, from the experimental point of view, the knowledge of atomic positions in the unit cell is critical for analysis of NMR/NQR results and quantum oscillation measurements that characterize the electronic band structure. Here we have reported a corrected structural model for $CePt_2In_7$, with different Pt-In stacking, that has a similar statistical agreement to the observed diffraction intensities as does the previously reported model (ref. 11). However, the absence of the significant number of Ce site vacancies, and strong Pt-In site disorder is much more consistent with reported physical properties. In addition, the Ce-Pt local coordination is much more consistent with that of the other members of the structural family. The density of states calculation reveals important differences between the two crystal structures. Instead of four two-dimensional bands at the Fermi level, calculations performed for the corrected structure reveal three bands that are more three dimensional. These results from the fact that in the corrected crystal structure the Ce-Pt distance is shorter, which allows for Ce-Pt electronic interactions and increases the electronic dimensionality. Finally, our DOS calculations show that the new structural model for $CePt_2In_7$ has a significantly lower energy than the one reported earlier.

**Acknowledgements**

The research performed at Gdansk University of Technology was financially supported by National Science Centre grant (DEC-2012/07/E/ST3/00584). The structural work performed at Princeton University was supported by the US Department of Energy, Division of Basic Energy Sciences, grant DOE-FG02-08-ER-46544.

**Figures**

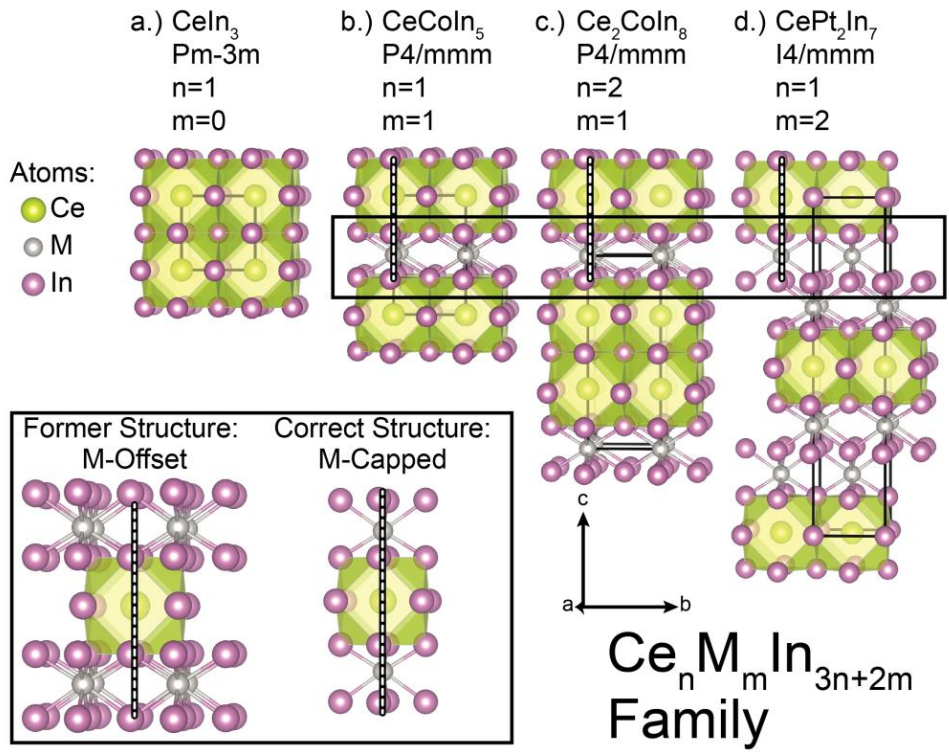

Figure 1. Crystal structures (tilted slightly from the *b-c* plane) of the Ce$_n$M$_m$In3$_{n+2m}$ family of compounds for n=1, m=0 (a), n=1, m=1 (b), n=2, m=1 (c), n=1 and m=2 (d). Green and purple balls represent Ce and In atoms, respectively. Grey balls represent transition metal atoms as described in the text. The correct structure model for CePt$_2$In$_7$ is shown. The area inside the box compares the Ce-Pt highlights the difference the former structural model and the one described here. The vertical dotted lines draw attention to the location of the Pt relative to Ce.

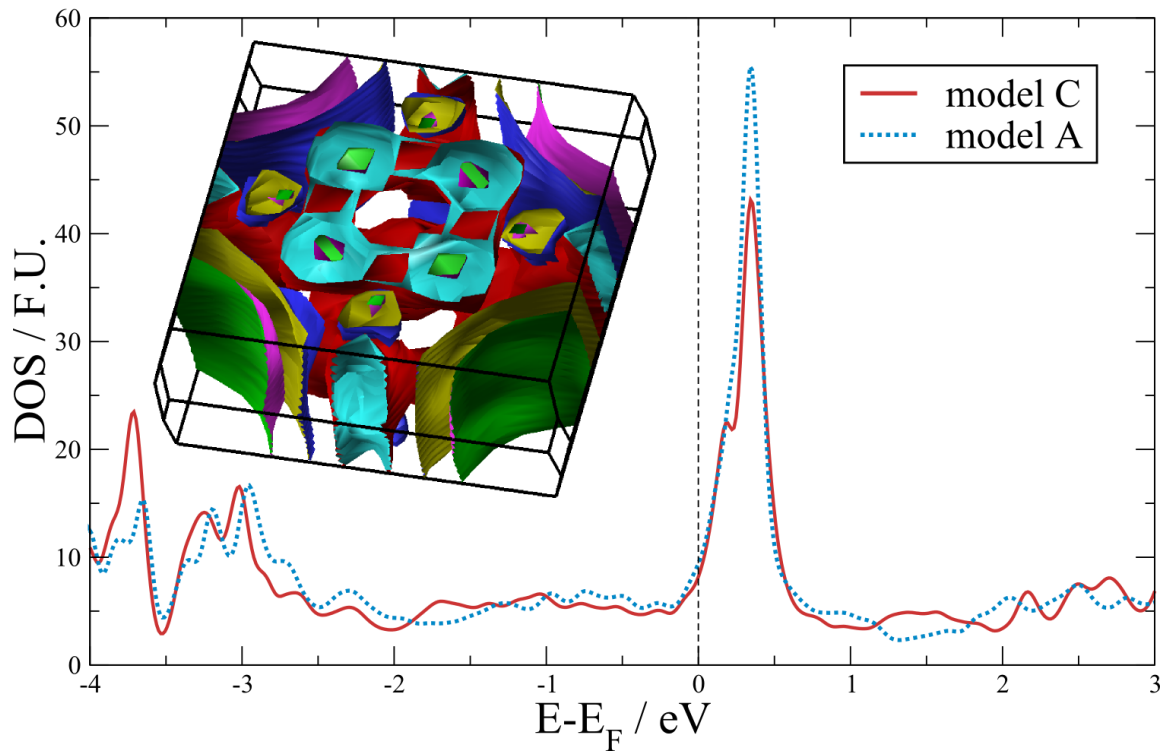

Figure 2: Density of states per formula unit for both structural models. Model (A) ordered version of former model, Model (C) the correct model. Inset: the Fermi-surface for the correct model.

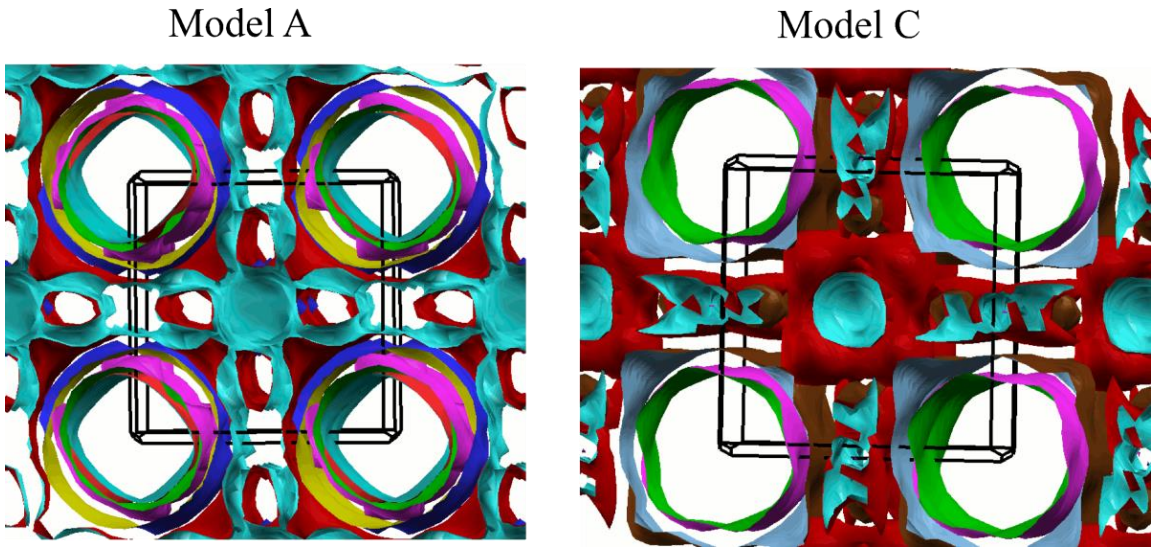

Figure 3. Comparison of the Fermi surfaces of the two models. The 1$^{st}$ Brillouin zone is drawn in black. In both cases, 4 bands cross the Fermi level. In the former model (A) there are 3 cylinders with little dispersion in the z-direction, whereas the Fermi surface becomes more 3-dimensional in the correct model (C) with only two 2-dimensional Fermi surfaces. One of the bands in the correct model is almost fully occupied and provides only barely visible small pockets and the zone boundary.

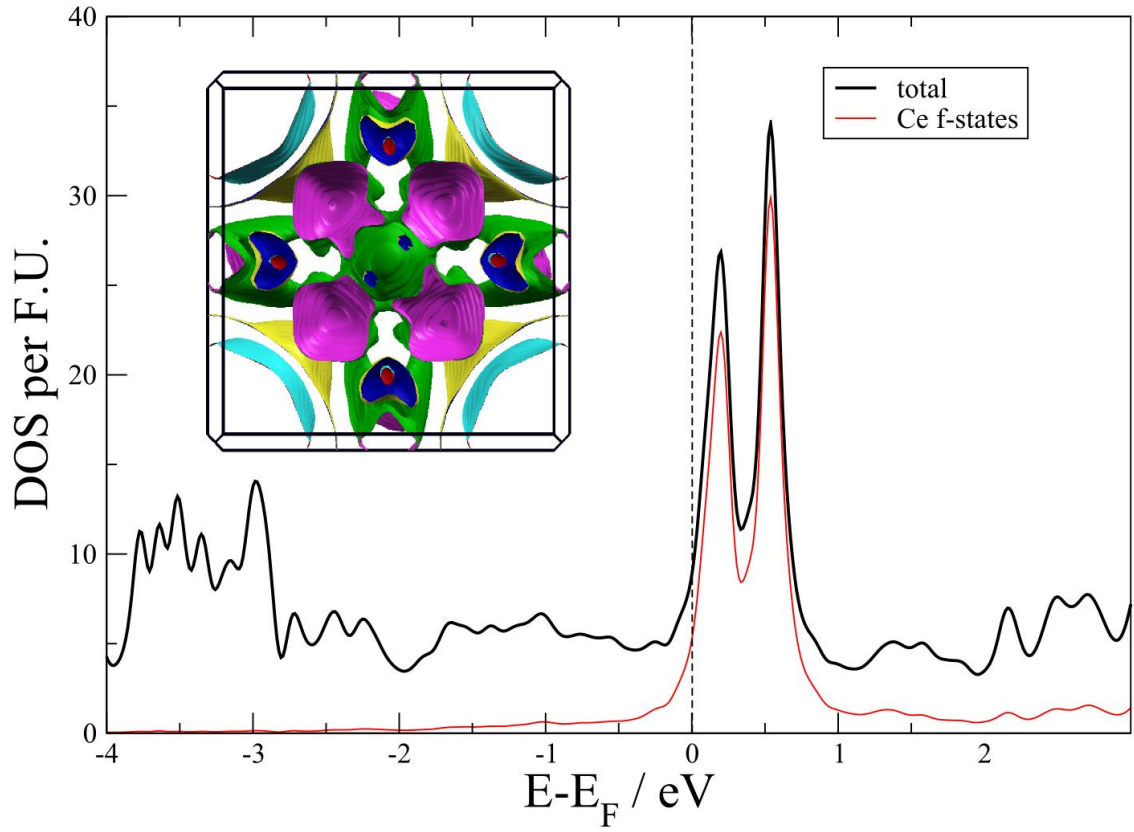

Figure 4. Density of states for the correct structural model (C) with spin-orbit coupling included. The DOS at the Fermi level is barely changed as is the Fermi surface. The Ce f-states are heavily influenced by spin-orbit coupling.

# Tables

Table 1. Single crystal x-ray data, refinement parameters and atomic coordinates for $CePt_2In_7$

|  | **Model A** | **Model B** (ref. 7, 11) | **Model C** (this work) |
|---|---|---|---|
| Empirical formula | $CePt_2In_7$ | $Ce_{0.85}Pt_{2.25}In_{6.75}$ | $CePt_2In_7$ |
| Formula weight | 83.38 | 83.32 | 83.38 |
| Temperature | 173(2) K | | |
| Wavelength | 7.1073 Å | | |
| Crystal system | Tetragonal | | |
| Space group | I 4/m m m | | |
| Unit cell dimensions | a = b = 4.5990(4) Å<br>c = 21.580(2) Å | | |
| Volume | 0.45644(9) $nm^3$ | | |
| Z | 32 | | |
| Density (calculated) | 9.707 $g/cm^3$ | 9.700 $g/cm^3$ | 9.707 $g/cm^3$ |
| Absorption coefficient | 52.629 $mm^{-1}$ | 55.095 $mm^{-1}$ | 52.629 $mm^{-1}$ |
| F(000) | 1114 | 1114 | 1114 |
| Crystal size | 0.008 x 0.045 x 0.055 $mm^3$ | | |
| Theta range for data collection | 3.777 to 38.544° | | |
| Index ranges | -7<=h<=7 -7<=k<=4, -37<=l<=34 | | |
| Reflections collected | 4607 | | |
| Independent reflections | 433 [R(int) = 0.0282] | | |
| Completeness to theta = 23.500° | 99.1 % | | |
| Refinement method | Full-matrix least-squares on $F^2$ | | |
| Data / restraints / parameters | 433 / 0 / 15 | 433 / 0 / 15 | 433 / 0 / 15 |
| Goodness-of-fit on $F^2$ | 2.143 | 1.089 | 1.022 |
| Final R indices [I>2sigma(I)] | R1 = 0.0450, wR2 = 0.1283 | R1 = 0.0280, wR2 = 0.0651 | R1 = 0.0266, wR2 = 0.0609 |
| R indices (all data) | R1 = 0.0458, wR2 = 0.1286 | R1 = 0.0286, wR2 = 0.0653 | R1 = 0.0273, wR2 = 0.0612 |
| Extinction coefficient | 0.0017(3) | 0.0013(1) | 0.0013(1) |
| Largest diff. peak and hole | 18.449 and -13.499 e.$Å^{-3}$ | 4.158 and -3.934 e.$Å^{-3}$ | 3.313 and -3.578 e.$Å^{-3}$ |

Table 2. Atomic coordinates and equivalent isotropic displacement parameters ($pm^2 \times 10^{-1}$) for each model as described in the text. U(eq) is defined as one third of the trace of the orthogonalized $U^{ij}$ tensor. Structure model C is the correct structure for $CePt_2In_7$

**Model A**

| Atom | Wyck. | x/a | y/b | z/c | Occ. | $U_{eq}$ |
|---|---|---|---|---|---|---|
| **Ce(1)** | 2a | 0 | 0 | 0 | 1 | 12(1) |
| **Pt(1)** | 4e | 0 | 0 | 0.3256(1) | 1 | 3(1) |
| **In(1)** | 2b | 0 | 0 | 0.5 | 1 | -3(1) |
| **In(2)** | 4d | 0 | 0.5 | 0.25 | 1 | 4(1) |
| **In(3)** | 8g | 0 | 0.5 | 0.1079(1) | 1 | 4(1) |

**Model B (ref. 7, 11)**

| Atom | Wyck. | x/a | y/b | z/c | Occ. | $U_{eq}$ |
|---|---|---|---|---|---|---|
| **Ce(1)** | 2a | 0 | 0 | 0 | 0.8499 | 6(1) |
| **Pt(1)** | 4e | 0 | 0 | 0.3256(1) | 1 | 3(1) |
| **Pt(2)** | 2b | 0 | 0 | 0.5 | 0.2512 | 2(1) |
| **In(1)** | 2b | 0 | 0 | 0.5 | 0.7488 | 2(1) |
| **In(2)** | 4d | 0 | 0.5 | 0.25 | 1 | 3(1) |
| **In(3)** | 8g | 0 | 0.5 | 0.1079(1) | 1 | 4(1) |

**Model C (this work)**

| Atom | Wyck. | x/a | y/b | z/c | Occ. | $U_{eq}$ |
|---|---|---|---|---|---|---|
| **Ce(1)** | 2b | 0 | 0 | 0.5 | 1 | 2(1) |
| **Pt(1)** | 4e | 0 | 0 | 0.3256(1) | 1 | 3(1) |
| **In(1)** | 2a | 0 | 0 | 0 | 1 | 4(1) |
| **In(2)** | 4d | 0 | 0.5 | 0.25 | 1 | 3(1) |
| **In(3)** | 8g | 0 | 0.5 | 0.1079(1) | 1 | 4(1) |

**References**

[1] F. Steglich, J. Aarts, C. D. Bredl, W. Lieke, D. Meschede, W. Franz, and H. Schäfer, Phys. Rev. Lett. **43**, 1892 (1979).

[2] K. Andres, J. E. Graebner, and H. R. Ott, Phys. Rev. Lett. **35**, 1779 (1975).

[3] C. Petrovic, P. G. Pagliuso, M. F. Hundley, R. Movshovich, J. L. Sarrao, J. D. Thompson, Z. Fisk and P Monthoux, J. Phys.: Condens. Matter **13**, L337 (2001).

[4] D. Kaczorowski, D. Gnida, A. P. Pikul, and V. H. Tran, Solid State Commun. **150**, 411 (2010).

[5] D. Kaczorowski, A. P. Pikul, D. Gnida, and V. H. Tran, Phys. Rev. Lett. **103**, 027003 (2009).

[6] K. Uhlirova, J. Prokleska, V. Sechovsky, and S. Danis, Intermetallics **18**, 2025 (2010).

[7] Zh.M. Kurenbaeva, E.V. Murashova, Y.D. Seropegin, H. Noël, A.I. Tursina, Intermetallics vol. **16**, 979 (2008).

[8] A. Tursina, S. Nesterenko, Y. Seropegin, H. Noel, D. Kaczorowski, J. Solid State Chem. **200**, 7 (2013).

[9] M. Kratochvilova, M. Dusek, K. Uhlirova, A. Rudajevova, J. Prokleska, B. Vondrackova, J. Custers, V. Sechovsky, J. of Crystal Growth **397**, 47 (2014).

[10] J.D. Thompson, Z. Fisk, J. Phys. Soc. Jpn. **81**, 011002 (2012).

[11] P.H. Tobash, F. Ronning, J.D. Thompson, B.L. Scott, P.J.W. Moll, B. Batlogg and E.D. Bauer, J. Phys.: Condens. Matter **24**, 015601 (2012).

[12] H. Sakai, Y. Tokunaga, S. Kambe, F. Ronning, E. D. Bauer, and J. D. Thompson, Phys. Rev. Lett. **112**, 206401 (2014).

[13] SADABS, Bruker (2007). Bruker AXS Inc., Madison, Wisconsin, USA

[14] SHELX-2013, Sheldrick, G. M. (2008). Acta Cryst. A64, 112--122.

[15] P. Blaha, K. Schwarz, G. Madsen, D. Kvasnicka, and J. Luitz, WIEN2K, An Augmented Plane Wave+Local Orbitals Program for Calculating Crystal Properties (Technische Universitat Wien,Vienna, Austria, 2001).

[16] J. P. Perdew, K. Burke, and M. Ernzerhof, Phys. Rev. Lett. **77**, 3865 (1996).

[17] E. D. Bauer, H. O. Lee, V. A. Sidorov, N. Kurita, K. Gofryk, J.-X. Zhu, F. Ronning, R. Movshovich, J. D. Thompson, and Tuson Park, Phys. Rev. B **81**,180507 (2010).